\documentclass[]{aastex631}

\usepackage[nolist]{acronym}

\begin{acronym}[AWGN]
\acro{ASKAP}{Australian Square Kilometre Array Pathfinder}
\acro{ARC}{Australian Research Council}
\acro{ATNF}{Australia Telescope National Facility}
\acro{CASDA}{CSIRO \ac{ASKAP} Science Data Archive}
\acro{CSIRO}{Australian Commonwealth Scientific and Industrial Research Organisation}
\acro{EMU}{Evolutionary Map of the Universe}
\acro{RMS}[rms]{root mean squared}
\acro{SN}{supernova}
\acro{SNR}{supernova remnant}
\acro{WISE}{Wide-Field Infrared Survey Explorer}
\end{acronym}

\newcommand{\g}{G308.73+1.38}
\newcommand{\ujybm}{\,$\mu$Jy\,beam$^{-1}$}
\def\arcmin{\hbox{$^\prime$}}
\newcommand{\D}{$^\circ$}

\begin{document}

\title{ASKAP$-$EMU Discovery of ``Raspberry'': a new Galactic SNR Candidate G308.73+1.38}
\correspondingauthor{s.lazarevic@westernsydney.edu.au}

\author[0000-0001-6109-8548]{Sanja Lazarevi\'c}
\affiliation{Western Sydney University, Locked Bag 1797, Penrith South DC, NSW 2751, Australia}
\affiliation{CSIRO Space and Astronomy, Australia Telescope National Facility, PO Box 76, Epping, NSW 1710, Australia}

\author[0000-0002-4990-9288]{Miroslav D. Filipovi\'c}
\affiliation{Western Sydney University, Locked Bag 1797, Penrith South DC, NSW 2751, Australia}

\author[0000-0003-4351-993X]{Bärbel S. Koribalski}
\affiliation{CSIRO Space and Astronomy, Australia Telescope National Facility, PO Box 76, Epping, NSW 1710, Australia}
\affiliation{Western Sydney University, Locked Bag 1797, Penrith South DC, NSW 2751, Australia}

\author[0009-0009-7061-0553]{Zachary J. Smeaton}
\affiliation{Western Sydney University, Locked Bag 1797, Penrith South DC, NSW 2751, Australia}

\author[0000-0002-6097-2747]{Andrew M. Hopkins}
\affiliation{School of Mathematical and Physical Sciences, 12 Wally’s Walk, Macquarie University, NSW 2109, Australia}

\author[0000-0001-5609-7372]{Rami Z. E. Alsaberi}
\affiliation{Western Sydney University, Locked Bag 1797, Penrith South DC, NSW 2751, Australia}

\author[0000-0002-0416-3267]{Velibor Velovi\'c}
\affiliation{Western Sydney University, Locked Bag 1797, Penrith South DC, NSW 2751, Australia}

\author[0009-0003-2088-9433]{Brianna D. Ball}
\affiliation{Department of Physics, University of Alberta, 4-181 CCIS, Edmonton, Alberta T6G 2E1, Canada}

\author[0000-0001-5953-0100]{Roland Kothes}
\affiliation{Dominion Radio Astrophysical Observatory, Herzberg Astronomy \& Astrophysics, National Research Council Canada, P.O. Box 248, Penticton, BC V2A 6J9, Canada}

\author[0000-0002-4814-958X]{Denis Leahy}
\affiliation{Department of Physics and Astronomy, University of Calgary, Calgary, Alberta, T2N 1N4, Canada}

\author[0000-0002-3137-473X]{Adriano Ingallinera}
\affiliation{INAF-Osservatorio Astrofisico di Catania, via S. Sofia 78, I-95123 Catania, Italy}

\begin{abstract}
We report the ASKAP discovery of a new Galactic \ac{SNR} candidate \g, which we name Raspberry. This new \ac{SNR} candidate has an angular size of 20\farcm7$\times$16\farcm7, and we measure a total integrated flux of 407$\pm$50\,mJy. We estimate Raspberry's most likely diameter of 10$-$30\,pc which would place it at a distance 3$-$5\,kpc, in the near side of the Milky Way's Scutum-Centaurus Arm. We also find a Stokes$-$V point source close to the centre of Raspberry with a $\sim$5$\sigma$ significance. This point source may be the remaining compact source, 
a neutron star, or possibly a pulsar,
formed during the initial supernova event. 
\end{abstract}


\section{Introduction} 
\label{sec:introduction}


Supernova remnants (SNRs) are the resultant expanding structures that remain after the death of massive stars in a supernova explosion. It is generally accepted that the population of known Galactic \acp{SNR} \cite[$\approx$300,][]{Green} vastly underrepresents the expected population~\citep{2023MNRAS.524.1396B} and the discovery of new Galactic \acp{SNR} is vital in filling this knowledge gap. 

\begin{figure*}[t!]
\centering
    \includegraphics[trim=0 0 0 0,width=\linewidth]{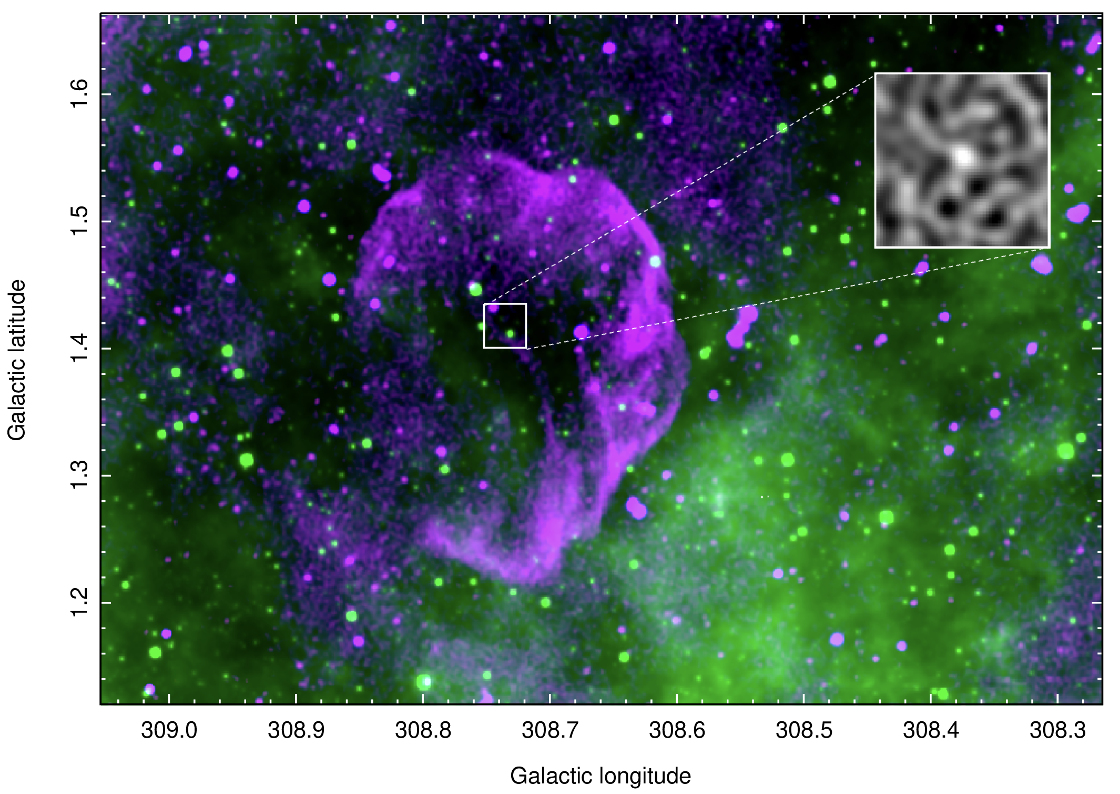}
    \caption{
    RGB composite image where the total intensity map of Raspberry, observed by \ac{ASKAP} at $\nu$\,=\,944\,MHz, is in red and blue while WISE~12\,$\mu$m infrared image is in green. To present the structure of Raspberry, we used different colourmaps and adjusted contrast levels. A linear scale is applied to all images. 
    The inset is the \ac{ASKAP} Stokes$-$V zoomed-in image showing the possible progenitor source.
    }
    \label{fig:1}
\end{figure*}

We serendipitously discovered a new Galactic \ac{SNR} candidate (Figure~\ref{fig:1}), dubbed Raspberry, in the Australian Square Kilometre Array Pathfinder~\citep[ASKAP,][]{2021PASA...38....9H} Evolutionary Map of the Universe survey~\citep[EMU,][]{Norris2021}. This discovery reinforces the ability of the latest generation radio sky surveys, such as EMU, to detect new low surface brightness sources, as previously demonstrated with SNR\,J0624-694~\citep{2022MNRAS.512..265F}, Ankora \citep[G288.8$-$6.3,][]{2023AJ....166..149F} 
and Diprotodon (G278.9+1.3, Filipovi\'{c} et al., in prep.). 

\section{Data}
\label{askap}

As a part of the large-scale ASKAP$-$EMU project, 
the SNR area of the sky was observed in December 2023 with a complete set of 36 ASKAP antennas at the central frequency of 943.4\,MHz and bandwidth of 288~MHz. All data are available through the CSIRO ASKAP Science Data Archive (CASDA\footnote{\url{https://research.csiro.au/casda}}). The observation containing this object is the tile EMU\_1342$-$60 corresponding to ASKAP scheduling block SB54095. The data was processed using the ASKAPsoft pipelines, including multi-frequency synthesis imaging, multi-scale clean, self-calibration and convolution to a common beam size \citep{2019ascl.soft12003G}. The resulting 943\,MHz EMU Stokes$-$I image has a \ac{RMS} sensitivity of $\sigma$\,=\,33\ujybm, while Stokes$-$V image \ac{RMS} is $\sigma$\,=\,20\ujybm. The synthesised beam for both images is 15\arcsec$\times$15\arcsec. 

\section{Results and Discussion} 
\label{sec:results}

We propose that Raspberry is a candidate for a new Galactic shell-type \ac{SNR} based primarily on its radio morphology (Figure~\ref{fig:1}).
The extended shell at radio frequencies seen in Raspberry resembles typical shell-type SNRs.
Additionally, we searched the Wide-Field Infrared Survey Explorer~\citep[WISE,][]{2010AJ....140.1868W} all-sky maps and found no infrared emission that spatially corresponds with the observed radio shell. This lack of infrared emission indicates that the radio shell is purely non-thermal, formed by synchrotron emission from ultra-relativistic particles energised by the expanding shock front, typical of a shell-type \ac{SNR} \citep{book2}.

We found that Raspberry is centred at RA(J2000)\,=\,13$^{\rm h}$39$^{\rm m}$25\fs5 and Dec(J2000)\,=\,$-$60\D56\arcmin29\farcs4 ($l$\,=\,308\fdg73 and $b$\,=\,+1\fdg38)
with
the angular size of 20\farcm7$\times$16\farcm7. The shell is composed of a filamentary structure with the brightest regions occurring on the western edge. We measure an integrated flux density of S$_I$\,=\,407$\pm$50\,mJy over the entire shell area.

The typical average value for an \ac{SNR} spectral index is $\alpha$\,=\,$-$0.5, obtained from theoretical models~\citep{Bell1978} and observations~\citep{2017ApJS..230....2B}. Assuming Raspberry's spectral index of $\alpha$\,=\,$-$0.5 and integrated flux density of S$_I$\,=\,407$\pm$50\,mJy, we calculate a scaled flux of S$_{1 \text{GHz}}$\,=\,389\,mJy giving surface brightness of $\Sigma_{1 \text{GHz}}$\,=\,1.4$\times$10$^{-20}$W\,m$^{-2}$\,Hz$^{-1}$. The $\Sigma$$-$$D$ calibration model from \cite{Pavlovic2018} works well for Galactic shell-type \acp{SNR} similar to Raspberry, and we compare our measured values to estimate the diameter. Comparing with~\citet[their Fig.~3]{Pavlovic2018}, we estimate a diameter of $D$\,=\,10$-$30\,pc. This diameter would place Raspberry at a distance of 3$-$5\,kpc, which is in the Milky Way's near side of the Scutum-Centaurus Arm, based on its Galactic longitude.

We detect a possible circularly polarised point source near the centre of Raspberry, observable only in the Stokes$-$V image (Figure~\ref{fig:1}, inset). The coordinates of this point source are RA(J2000)\,=\,13$^{\rm h}$39$^{\rm m}$26\fs8 and Dec(J2000)\,=\,$-$60\D54\arcmin38\farcs5, giving the source an offset of 111\arcsec\ from Raspberry's geometric centre. The point source has a Stokes$-$V flux density of 100\,$\mu$Jy. Given that the local \ac{RMS} in the Stokes$-$V image is 20\,$\mu$Jy beam$^{-1}$, we estimate that the point source has a significance of $\sim$5$\sigma$. However, the source is not detected in the Stokes$-$I image. This point source could potentially be Raspberry's compact source, 
a neutron star, or possibly a pulsar,
produced in the initial \ac{SN} event. No optical or infrared counterpart can be identified in existing surveys.

\section{Conclusion}
We present the serendipitous discovery of a new Galactic \ac{SNR} candidate Raspberry (\g). The \ac{SNR} identification is primarily based on the object's radio-continuum morphology and the absence of an infrared counterpart. We observe a Stokes$-$V point source close to Raspberry's geometrical centre that has the potential to be a central compact source of the \ac{SNR}. 

Further multi-frequency observations are required to confirm Raspberry's identity as an \ac{SNR}. Particularly, additional radio-continuum bands would allow us to confirm the non-thermal origin of the radio emission (via spectral index); full Stokes parameters would allow us to study polarisation which is a strong characteristic of the shell SNRs; X-ray and high-energy observations would allow us to further determine the multi-frequency spectrum of this object; and a dedicated pulsar search would allow us to determine the true nature of the potential point source at the heart of Raspberry.

\section*{Acknowledgments}
\begin{acknowledgments}
This scientific work uses data obtained from Inyarrimanha Ilgari Bundara/the Murchison Radio-astronomy Observatory. We acknowledge the Wajarri Yamaji People as the Traditional Owners and native title holders of the Observatory site. CSIRO’s ASKAP radio telescope is part of the Australia Telescope National Facility (https://ror.org/05qajvd42). Operation of ASKAP is funded by the Australian Government with support from the National Collaborative Research Infrastructure Strategy. ASKAP uses the resources of the Pawsey Supercomputing Research Centre. Establishment of ASKAP, Inyarrimanha Ilgari Bundara, the CSIRO Murchison Radio-astronomy Observatory and the Pawsey Supercomputing Research Centre are initiatives of the Australian Government, with support from the Government of Western Australia and the Science and Industry Endowment Fund.
\end{acknowledgments}

\bibliography{raspberry}{}
\bibliographystyle{aasjournal}

\end{document}